\documentclass[a4paper,11pt]{article}
\pdfoutput=1 

\usepackage{jcappub} 

\usepackage[T1]{fontenc} 
\usepackage{amsmath}
\usepackage{scalerel}
\usepackage{tensor}
\usepackage{color, colortbl}
\definecolor{light-gray}{gray}{0.95}
\definecolor{darkred}{rgb}{0.7,0,0}

\title{\boldmath How attractive is the isotropic attractor solution of axion-SU(2) inflation?}

\author[\dag,1]{Ira Wolfson,\note{Corresponding author.}}
\author[\dag]{Azadeh Maleknejad,}
\author[\dag\ast]{and Eiichiro Komatsu}

\affiliation[\dag]{Max Planck Institute for Astrophysics,\\ Karl-Schwarzschild-Str. 1, 85748 Garching, Germany
}
\affiliation[\ast]{Kavli Institute for the Physics and Mathematics of the Universe (Kavli IPMU, WPI),\\ UTIAS, The University of Tokyo, Chiba, 277-8583, Japan
}

\emailAdd{irawolf@mpa-garching.mpg.de}
\emailAdd{amalek@mpa-garching.mpg.de}
\emailAdd{komatsu@mpa-garching.mpg.de}

\abstract{
The key to the phenomenological success of inflation models with axion and SU(2) gauge fields is the isotropic background of the SU(2) field. Previous studies showed that this isotropic background is an attractor solution during inflation starting from anisotropic (Bianchi type I) spacetime; however, not all possible initial anisotropic parameter space was explored. In this paper, we explore more generic initial conditions without assuming the initial slow-roll dynamics. We find some initial anisotropic parameter space which does {\it not} lead to the isotropic background, but to violation of slow-roll conditions, terminating inflation prematurely. The basin of attraction increases when we introduce another scalar field acting as inflaton and make the axion-SU(2) system a spectator sector. Therefore, the spectator axion-SU(2) model is phenomenologically more attractive.}

\begin{document}
\maketitle
\flushbottom
\section{Introduction}
Cosmic inflation \cite{Guth:1980zm,Sato:1980yn} based on a single slowly rolling scalar field \cite{Linde:1981mu,Albrecht:1982wi,Linde:1983gd} yields primordial scalar perturbations which are nearly scale-invariant and nearly Gaussian \cite{Mukhanov:1981xt,Starobinsky:1982ee,Hawking:1982cz,Bardeen:1983qw}. It can also yield primordial tensor perturbations that are nearly scale invariant and nearly Gaussian \cite{Starobinsky:1979ty}.  

The presence of gauge fields during inflation changes this picture, as they can produce non-scale invariant, non-Gaussian scalar and tensor perturbations. For example, a U(1) gauge field coupled to an axion sources scalar and tensor perturbations at quadratic order, resulting in highly non-Gaussian perturbations  \cite{Anber:2009ua,Barnaby:2010vf, Barnaby:2011vw,Anber:2012du}. The lack of detection of scalar non-Gaussianities \cite{Akrami:2019izv} rules out a large parameter space of such models. Nevertheless one new phenomenology predicted by these models is that tensor perturbations become chiral \cite{Barnaby:2012xt,Namba:2015gja,Anber:2012du}. 

U(1) gauge field models face an additional challenge as they produce a preferred direction in the universe, breaking the statistical isotropy of perturbations (see \cite{Maleknejad:2012fw} for a review), that is not detected \cite{Kim:2013gka,Akrami:2018odb}. 

Models with SU(2) gauge fields coupled with an axion avoid both challenges. Since the SU(2) gauge algebra is isomorphic to the SO(3) algebra, the rotational invariance is protected and the statistical isotropy remains unbroken \cite{Maleknejad:2011jw,Maleknejad:2011sq}. As for perturbations there is a large parameter space in which SU(2) gauge fields produce chiral tensor perturbations with negligible contributions to scalar perturbations at the linear order \cite{Dimastrogiovanni:2012ew,Adshead:2013qp,Maleknejad:2012fw}. See \cite{Papageorgiou:2018rfx,Papageorgiou:2019ecb} for study of non-linear perturbations.

The homogeneous background of gauge fields leads to particle production that may produce significant back-reaction on the dynamics of the homogeneous gauge and scalar fields. The back-reaction due to charged scalar fields and fermions coupled to SU(2) gauge fields and spin-2 particles of SU(2) further restricts the phenomenologically viable parameter space of these models \cite{Lozanov:2018kpk,Mirzagholi:2019jeb,Maleknejad:2018nxz,Maleknejad:2019hdr,Domcke:2018eki,Domcke:2018gfr,Domcke:2019qmm}.

The key to the phenomenological success of these models is the isotropic background of the SU(2) gauge fields. Given an anisotropic background initial geometry, how did the isotropic background emerge? Previous studies have shown that the isotropic background of the SU(2) gauge field is an attractor solution \cite{Maleknejad:2013npa,Adshead:2012kp}. However not all possible initial configurations lead to an isotropic end state, defying the so-called cosmic 'no-hair' conjecture \cite{Maleknejad:2011jr,Adshead:2018emn}. Moreover, as we show in this paper, some lead to violation of slow-roll conditions terminating inflation prematurely. We also show that the attractor's basin of attraction increases when we introduce another scalar field acting as inflaton and demote the axion-SU(2) system to a spectator sector. Thus the axion-SU(2) spectator model \cite{Dimastrogiovanni:2016fuu} is both phenomenologically rich and observationally viable.

This paper is organized as follows. In section \ref{sec:Isotropic Background with SU(2)} we review SU(2) gauge fields in an isotropic background Friedmann-Robertson-Walker (FRW) metric, coupled to axion fields via a Chern-Simons (CS) topological term. Section \ref{sec:Anisotropic} introduces the chromo-natural (CN) system embedded in an anisotropic Bianchi type I metric. We go through some of the theoretic consequences and present new results of numerical analysis. In section \ref{sec:Anisotropic-Spectator} the CN system is demoted to a spectator status, and we analytically and numerically study this scalar-driven inflation with a spectator axion-SU(2) component. Our numerical study is focused on the initial condition basin of the isotropic attractor solution. Finally we conclude in section \ref{sec:Conclusion}.\\\\
We work in the mostly positive signature ($-,+,+,+$), and in natural units where $c=\hbar=1$. The reduced Planck mass $M_{Pl}=\sqrt{\frac{1}{8\pi G}}$ is set to $1$.
\section{Isotropic Background of the SU(2) gauge field}\label{sec:Isotropic Background with SU(2)}
\subsection{Axion-SU(2) inflation}
The axion-SU(2) inflation is a class of inflationary models where SU(2) gauge fields contribute to the physics of inflation. Within this family of models we look at the chromo-natural (CN) model in which we have an axion field with a cosine potential. The action is given by \cite{Adshead:2012kp}:
\begin{align}
    S=\int d^4 x\sqrt{-g}\left[-\frac{R}{2} -\frac{1}{4}F^a_{\mu\nu}F_a^{\mu\nu} -\frac{1}{2}\left(\partial_{\mu}\chi\right)^2 -\mu^4\left(1+\cos{\frac{\chi}{f}}\right)+\frac{\lambda \chi}{8f} \Tilde{F}^a_{\mu\nu}F_a^{\mu\nu}\right],\label{eq:CN_action}
\end{align}
where $\chi$ is the axion field with the potential $V\left(\chi\right)=\mu^4\left(1+\cos \frac{\chi}{f}\right)$, $f$ is the axion decay constant, $F^a_{\mu\nu}$ is the SU(2) field strength tensor:
\begin{align}
    F^a_{\mu\nu}=\partial_{\mu}A^a_{\nu}-\partial_{\nu}A^a_{\mu}-g_{\scaleto{A}{4pt}}\epsilon^a_{bc}A^b_{\mu}A^c_{\nu}\; ,
\end{align}
$\Tilde{F}^a_{\mu\nu}$ is its Hodge dual, $g_{\scaleto{A}{4pt}}$ is the gauge coupling, and $\epsilon^{a}_{bc}$ is the non-Abelian algebra's set of structure constants.
The CN model can be embedded in a simple FRW metric, as was done in \cite{Maleknejad:2011sq, Adshead:2012kp,Obata:2014loa,Adshead:2016omu} or in some other metric (see \cite{Maleknejad:2013npa} for Bianchi type I case).
The coupling of the axion and gauge fields through the CS term enables the kinetic gauge-field term to exchange energy with the axion potential, effectively augmenting the conventional drag term for the axion and prolonging slow-roll.\\\\
When embedded in an FRW metric, the gauge fields contribute to the energy density and supports the isotropic inflation of space. This can be seen in the Friedmann equations: 
\begin{align}
\left\{\begin{array}{cc}
     H^2=&\frac{\rho_{\scaleto{A}{4pt}} +\rho_{\chi}}{3}\\
     &\\
     \frac{\ddot{a}}{a}=&-\frac{\rho_{\scaleto{A}{4pt}} +\rho_{\chi}+3P_{\scaleto{A}{4pt}} +3P_{\chi}}{6}
\end{array}\right.\; ,
\end{align}
with the energy densities given by:
\begin{align}
    \rho_{\chi}=\frac{\dot{\chi}^2}{2}+\mu^4\left(1+\cos\frac{\chi}{f}\right),\\
    \rho_{\scaleto{A}{4pt}}=2\frac{\delta \mathcal{L_{\scaleto{A}{4pt}}}}{\delta F^{a\; 0}_{\;\sigma}} F^a_{\sigma 0}-\mathcal{L_{\scaleto{A}{4pt}}},
\end{align}
where we call the term:
\begin{align}
   \mathcal{L_{\scaleto{A}{4pt}}}= -\frac{1}{4}F^a_{\mu\nu}F_{a}^{\mu\nu} +\frac{1}{4} \left(\frac{\lambda\chi}{2f}\right)\Tilde{F}^a_{\mu\nu}F_a^{\mu\nu}
\end{align}
the gauge fields ($A$) Lagrangian. The pressure contributions are given by:
\begin{align}
    P_{\chi}=\dot{\chi}^2-\rho_{\chi}\;,\\
    \nonumber \\
    P_{\scaleto{A}{4pt}}=\frac{\rho_{\scaleto{A}{4pt}}}{3} \; .
\end{align}
\\
In the FRW spacetime 
\begin{align}
    ds^2=-dt^2 +a(t)^2\left(d{\bf x}^2\right)\;,
\end{align}
the CN system has an isotropic and homogeneous background solution. In particular, the SU(2) gauge field in the temporal gauge has the following solution \cite{Maleknejad:2011jw,Maleknejad:2011sq}:
\begin{align}
    A^a_0=0,\hspace{15pt} A^a_i=\psi(t)a(t)\delta^a_i \; , \label{eq:Ansatz-FRW}
\end{align}
where $\psi(t)$ is a pseudo-scalar that parametrizes the effective field value of the gauge field. The energy density for the gauge fields then become:
\begin{align}
    \rho_{\scaleto{A}{4pt}}=\frac{3}{2}\left(\dot{\psi}+H\psi\right)^2+\frac{3g_{\scaleto{A}{4pt}}^2\psi^4}{2},
\end{align}
and equations of motion are given by \cite{Adshead:2012kp}:
\begin{align}
   \ddot{\chi}+3H\dot{\chi}=\frac{\mu^4}{f}\sin{\frac{\chi}{f}}-3g_{\scaleto{A}{4pt}}\frac{\lambda}{f}\psi^2\left(\dot{\psi}+H\psi\right)\; , \\
   \nonumber \\
   \ddot{\psi}+3H\dot{\psi}=-(2H^2+\dot{H})\psi-2g_{\scaleto{A}{4pt}}^2\psi^3 +g_{\scaleto{A}{4pt}}\frac{\lambda}{f}\psi^2 \dot{\chi}\; .
\end{align}
Asserting slow-roll of all fields, we obtain a set of master equations:
\begin{align}
    \dot{\chi}\simeq\frac{\frac{\mu^4}{f}\sin{\frac{\chi}{f}}-3g_{\scaleto{A}{4pt}}\frac{\lambda}{f}\psi^2\left(\dot{\psi}+H\psi\right) }{3H}\; , \label{eq:chi_master}\\
   \nonumber \\
   \dot{\psi}\simeq\frac{-2H^2\psi-2g_{\scaleto{A}{4pt}}^2\psi^3 +g_{\scaleto{A}{4pt}}\frac{\lambda}{f}\psi^2 \dot{\chi}}{3H}\label{eq:psi_master}\; .
\end{align}
Furthermore, discarding $\dot{\chi}$ and $\dot{\psi}$ as minuscule gives \cite{Adshead:2012kp}:
\begin{align}
    \psi\big|_{FRW}\simeq \left( \frac{\mu^4\sin{\frac{\chi}{f}}}{3g_{\scaleto{A}{4pt}}H\lambda}\right)^{\frac{1}{3}}\; .\label{eq:psi_FRW}
\end{align}
As the axion kinetic term is never important in this setup, The slow-roll parameters are approximated by:
\begin{align}
    \varepsilon_{\scaleto{H}{4pt}}\equiv -\frac{\dot{H}}{H^2}\simeq \frac{3g_{\scaleto{A}{4pt}}^2\psi^4}{H^2}+\psi^2 \simeq \frac{3g_{\scaleto{A}{4pt}}^2\psi^4}{\mu^4\left(1+\cos{\frac{\chi}{f}}\right)}+\psi^2,\label{eps_adshead}
\end{align}
which holds when the energy density is dominated by the axion potential (i.e. slow-roll), and $\eta_{\scaleto{H}{4pt}}$ is given by:\footnote{Note that in \cite{Adshead:2012kp}  $\eta_{\scaleto{H}{2pt}}\equiv\frac{d\ln \varepsilon_{\scaleto{H}{2pt}}}{dN}$ whereas we define $\eta_{\scaleto{H}{2pt}}\equiv \frac{\ddot{H}}{H\dot{H}}$.}
\begin{align}
    \eta_{\scaleto{H}{4pt}}\equiv \frac{\ddot{H}}{H\dot{H}} \simeq 2g_{\scaleto{A}{4pt}}^2\frac{\psi^4}{H^2}+\frac{\dot{\psi}}{H\varepsilon_{\scaleto{H}{4pt}}}\left(12g_{\scaleto{A}{4pt}}^2\frac{\psi^3}{H^2}+2\psi\right)-2\varepsilon_{\scaleto{H}{4pt}}.\label{eta_adshead}
\end{align}
In \cite{Adshead:2012kp} the authors point out that the second term in \ref{eta_adshead} has to be of $\mathcal{O}(\varepsilon^2)$ to facilitate slow-roll inflation. Additionally, \ref{eps_adshead} and \ref{eta_adshead} suggest (for $\varepsilon_H\ll 1$ and $\eta_H=\mathcal{O}(\varepsilon_H^2)$) 
\begin{align}
    \psi^2\ll\frac{H}{\sqrt{2}g_{\scaleto{A}{4pt}}},\hspace{15pt} |\psi|\ll \frac{\mu}{\sqrt{g_{\scaleto{A}{4pt}}}},
\end{align}
which place strict conditions on $\psi$ given the typical value of $\mu$.
Embedded in an FRW metric, the gauge fields are isotropic, and to leading order homogeneous.
\subsection{Spectator Axion-SU(2) Sector}\label{sec:Spectator_FRW} 
Relegating the CN system to a spectator status by adding a dominant minimally coupled scalar inflaton component, the action becomes \cite{Dimastrogiovanni:2016fuu}:
\begin{align}
    S=\int d^4 x \sqrt{-g}\left[-\frac{R}{2} -\frac{1}{4}F^a_{\mu\nu}F_a^{\mu\nu} -\frac{\left(\partial_{\mu}\chi\right)^2}{2} -\mu^4\left(1+\cos{\frac{\chi}{f}}\right)+\frac{\lambda \chi}{8f} \Tilde{F}^a_{\mu\nu}F_a^{\mu\nu}-\frac{\left(\partial_{\mu}\phi\right)^2}{2}-V(\phi)\right],\label{eq:CNspect_action}
\end{align}
which is shorthanded as:
\begin{align}
    S=S_{CN}+S_{\phi},
\end{align}
where the subscript $CN$ stands for the chromo-natural action given in Eq. \ref{eq:CN_action}.\\
Due to the minimal coupling between the axion-SU(2) sector and the inflaton, the equations of motion for the axion and SU(2) fields are the same and using the same ansatz as in \ref{eq:Ansatz-FRW} the matter Lagrangian can be read as:
\begin{align}
    \mathcal{L}_{m}=\frac{3}{2}\left(\frac{1}{a^2}\left(\frac{\partial(a\psi)}{\partial t}\right)^2 -g_{\scaleto{A}{4pt}}^2\psi^4\right)+\frac{\dot{\chi}^2}{2} -\mu^4\left(1+\cos{\frac{\chi}{f}}\right) -3g_{\scaleto{A}{4pt}}\frac{\lambda\chi}{f}\frac{\psi^2}{a}\frac{\partial(a\psi)}{\partial t}+\frac{\dot{\phi}^2}{2}-V(\phi)\; .
\end{align}
Since the metric is still FRW, the Friedmann equations remain the same:
\begin{align}
    H^2=&\frac{\rho}{3}\\ \nonumber
    \\
    \frac{\ddot{a}}{a}=&-\frac{\rho+3P}{6},
\end{align}
and the total energy density is given by
\begin{align}
    \rho=\rho_{\chi}+\rho_{\scaleto{A}{4pt}}+\rho_{\phi},
\end{align}
in which $\rho_{\chi}$ and $\rho_{\scaleto{A}{4pt}}$ are the same as before, and $\rho_{\phi}$ is: 
\begin{align}
    \rho_{\phi}=\frac{\dot{\phi}^2}{2}+V(\phi).
\end{align}
The pressure is also linearly additive, so the total pressure is given by:
\begin{align}
    P=P_{\chi}+P_{\scaleto{A}{4pt}}+P_{\phi},\label{eq:Spectator pressure}
\end{align}
with
\begin{align}
    P_{\phi}=\frac{1}{2}\dot{\phi}^2 -V(\phi).
\end{align}
The slow-roll parameters are given by:
\begin{align}
    \varepsilon_H\equiv\frac{-\dot{H}}{H^2}=\frac{3}{2}\left(\frac{P+\rho}{\rho}\right),
\end{align}
and
\begin{align}
    \eta_H\equiv\frac{\ddot{H}}{H\dot{H}}=\sqrt{\frac{3}{\rho}}\frac{\partial}{\partial t}\ln{\left(|\rho+P|\right)}.
\end{align}
Taking the inflaton pressure and energy as dominant we can write these as:
\begin{align}
    P=P_{\phi}+\Delta P;\hspace{15pt} \rho=\rho_{\phi}+\Delta\rho,
\end{align}
where we imply $\frac{\Delta \rho}{\rho_{\phi}},\frac{\Delta P}{P_{\phi}}\ll 1$.
In this approximation the quantity $\frac{P+\rho}{\rho}$ becomes:
\begin{align}
    \frac{P+\rho}{\rho}\simeq\frac{P_{\phi}+\rho_{\phi}}{\rho_{\phi}}+\frac{1}{\rho_{\phi}}\left[\Delta\rho +\Delta P\right] -\frac{\Delta \rho}{\rho_{\phi}}\left[\frac{P_{\phi}+\rho_{\phi}}{\rho_{\phi}}+\frac{1}{\rho_{\phi}}\left[\Delta\rho +\Delta P\right]\right],
\end{align}
in which by virtue of the SU(2)-axion being a spectator sector, the third term is suppressed.\\
This yields:
\begin{align}
    \varepsilon\simeq\varepsilon_{\phi}+\frac{3}{2\rho_{\phi}}\left[P_{\chi}+P_{\scaleto{A}{4pt}}+\rho_{\chi}+\rho_{\scaleto{A}{4pt}}\right]=\varepsilon_{\phi}+\varepsilon_{\scaleto{CN}{4pt}}\;,
\end{align}
where $\varepsilon_{\phi}\equiv\frac{3(P_{\phi}+\rho_{\phi})}{2\rho_{\phi}}$ and $\varepsilon_{\scaleto{CN}{4pt}}\equiv\frac{3(P_{\chi}+P_{\scaleto{A}{4pt}}+\rho_{\chi}+\rho_{\scaleto{A}{4pt}})}{2\rho_{\phi}}$. From here on we dispose of the subscript $H$ on $\varepsilon$ and $\eta$. If we want to compare the slow-roll parameter of a spectator system and the CN system this can be written as:
\begin{align}
\varepsilon_{\text{spectator}}\simeq\varepsilon_{\phi}+\frac{\rho_{\scaleto{CN_0}{4pt}}}{\rho_\phi}\varepsilon_{\scaleto{CN_0}{4pt}},    
\end{align}
where the subscript $CN_0$ denotes the original CN system, without the inflaton potential, as in Eq.~\eqref{eps_adshead}.\\\\
The second slow-roll parameter expansion has two limits. In the limit of $\varepsilon_{\phi}\ll\varepsilon_{\scaleto{CN}{4pt}}$ we have:
\begin{align}
    \eta_{\text{spectator}}=\eta_{\phi}+\sqrt{\frac{3}{\rho_{\phi}}}\frac{\partial}{\partial t}\left(\frac{\varepsilon_{\scaleto{CN}{4pt}}}{\varepsilon_{\phi}}\right),\label{eq:FRW_CN_eta}
\end{align}
where $\eta_{\phi}\equiv \sqrt{\frac{3}{\rho_{\phi}}}\frac{\partial}{\partial t}\ln(|\rho_{\phi}+P_{\phi}|)$, the second term is much smaller than the first. Thus if we assume a well behaved Hubble parameter over the inflationary period, we can safely ignore that term. Therefore the second slow-roll parameter is also dominated by the scalar-field-associated parameter. The only caveat for this assumption is that we have to make sure $\varepsilon_{\scaleto{CN}{4pt}}$ does not oscillate too much; otherwise, the temporal derivative might dominate Eq.~\ref{eq:FRW_CN_eta}.\\
In the limit of $\varepsilon_{\scaleto{CN}{4pt}}\ll\varepsilon_{\phi}$ the term for $\eta_{\text{spectator}}$ becomes: 
\begin{align}
    \eta_{\text{spectator}}\simeq \eta_{\scaleto{CN}{4pt}}+\sqrt{\frac{3}{\rho_{\phi}}}\frac{\partial}{\partial t}\left(\frac{\varepsilon_{\phi}}{\varepsilon_{\scaleto{CN}{4pt}}}\right),\label{eq:eta_BIG_CN}
\end{align}
where $\eta_{\scaleto{CN}{4pt}}\equiv \sqrt{\frac{3}{\rho_{\phi}}}\partial_t\ln\left(|\rho_{\scaleto{CN}{4pt}}+P_{\scaleto{CN}{4pt}}|\right)$. However, we can construct $\dot{\varepsilon}_{\phi}$ to be sufficiently small such that:
\begin{align}
    \eta_{\text{spectator}}\simeq \eta_{\scaleto{CN}{4pt}}.
\end{align}
These expressions for $\eta_{\text{spectator}}$ suggest that in areas where the CN system is unstable, care should be taken choosing the inflaton potential. 
\section{Axion-SU(2) system in a Bianchi Type I spacetime}\label{sec:Anisotropic}
So far, we have discussed axion-SU(2) systems embedded in an FRW metric. We now turn to consider a homogeneous but anisotropic background geometry. Here we assume an axisymmetric Bianchi type I metric:
\begin{align}
    ds^2=-dt^2 +e^{2\alpha(t)}\left[e^{-4\sigma(t)}dx^2 +e^{2\sigma(t)}\left(dy^2+dz^2\right)\right],\label{geometry:BtI}
\end{align}
in which $e^{\alpha(t)}$ is the isotropic scale factor, $a(t)$, and the metric anisotropy is represented by $e^{\sigma(t)}$. As with the case of isotropic background, we define the isotropic Hubble parameter as:
\begin{align}
    H\equiv\dot{\alpha}.
\end{align} 
When $\sigma(t)=C$ for some constant $C$, the axes can always be rescaled such that the metric reduces to the FRW one. In other words, only the time varying part of $\sigma$ is a physical quantity.\\
The spatial triads now have the following form:
\begin{align}
    e^a_1=e^{\alpha-2\sigma}\delta^a_1,\quad e^a_2=e^{\alpha+\sigma}\delta^a_2,\quad \text{and}\quad e^a_3=e^{\alpha+\sigma}\delta^a_3\;.
\end{align}
In this geometry the energy-momentum tensor has the diagonal form of:
\begin{align}
    T^{\mu}_{\nu}=\left(\begin{array}{cccc}
        -\rho(t) & & &  \\
         &P(t)-2\Tilde{P}(t) & & \\
         & & P(t)+\Tilde{P}(t)& \\
         & & & P(t)+\Tilde{P}(t)\\
    \end{array}\right),
\end{align}
where $\rho$ is the energy density, $P$ is the isotropic pressure and $\Tilde{P}$ is the anisotropic pressure which parametrizes the amount of anisotropy in the energy-momentum tensor. The dynamics in Bianchi cosmology is specified by the following field equations:
\begin{align}
    \dot{\alpha}^2 -\dot{\sigma}^2=\frac{\rho}{3}, \label{Friedmann}
\end{align}
\begin{align}
    \ddot{\sigma}+3\dot{\alpha}\dot{\sigma}=\frac{\Tilde{P}}{3},\label{Anisotropic_PART}
\end{align}
\begin{align}
    \ddot{\alpha}+3\dot{\sigma}^2=-\frac{\rho +P}{2}.\label{2nd_friedmann}
\end{align}
From \ref{Anisotropic_PART} we realize that $\Tilde{P}$ is the source of the anisotropy $\dot{\sigma}$. In its absence the initial anisotropy is exponentially suppressed, at a time scale $\dot{\alpha}^{-1}$. However, a non-zero anisotropic pressure can lead to non-trivial dynamics of spatial anisotropies in the metric. When $\dot{\alpha}\equiv H(t)$ and $\sigma=C$, the equations above revert to the usual FRW Friedmann equations. \\
We follow the formalism of \cite{Maleknejad:2013npa} to simplify analytical treatment of the model. Choosing the temporal gauge the consistent truncation for the gauge fields gives:
\begin{align}
    A^{a}_{\mu}=\left\{\begin{array}{lr}
         0&\mu=0  \\
         \psi_i e^{a}_i& \mu=i  
    \end{array}\right.\; .
\end{align}
The axial symmetry of the metric in the $y-z$ plane yields $\psi_2=\psi_3$; thus, the explicit form of our ansatz is given by:
\begin{align}
    A^{a}_i=diag(e^{\alpha-2\sigma}\psi_1,e^{\alpha+\sigma}\psi_2,e^{\alpha+\sigma}\psi_3).
\end{align}
We now introduce the following field re-definitions:
\begin{align}
   \psi_1(t) =\frac{\psi(t)}{\beta^2(t)};\quad \psi_2(t)=\psi_3(t)=\beta(t)\psi(t),
\end{align}
in which $\psi(t)$ represents the {\it isotropic} field and $\beta(t)$ parametrizes {\it anisotropy} in the gauge fields, with $\beta=\pm 1$ being the isotropic gauge field configuration. Moreover, as the point $\beta=0$ is a singularity, $\beta$ cannot change sign.\\ With this reformulation the CN matter Lagrangian from the action \ref{eq:CN_action} now reads:
\begin{align}
   \mathcal{L}_m&=\beta^2 \left(\psi \left(\dot{\sigma}+\dot{\alpha}+\frac{\dot{\beta}}{\beta}\right)+\dot{\psi}\right)^2+\frac{1}{2 \beta^4}\left(\psi\left(-2 \dot{\sigma}+\dot{\alpha}-\frac{2 \dot{\beta}}{\beta}\right)+\dot{\psi }\right)^2\\ \nonumber
   &-\frac{3 g_{\scaleto{A}{4pt}} \lambda  \chi \psi^2 \left(\psi\dot{\alpha}+\dot{\psi}\right)}{f}-\mu^4 \left(1+\cos\frac{\chi }{f}\right)-\frac{1}{2} g_{\scaleto{A}{4pt}}^2 \beta^4 \psi^4-\frac{g_{\scaleto{A}{4pt}}^2 \psi^4}{\beta^2}+\frac{1}{2} \dot{\chi}^2.
\end{align}
The field equation of the axion is only coupled to the isotropic part of the gauge field, $\psi$. Therefore, the axion field equation remains unchanged and similar to Eq.~\ref{eq:chi_master}. It implies that only the isotropic gauge configuration is sourced by the axion. Later we see that it leads to the exponential decay of the anisotropic part by inflation.\\
The energy density $\rho$ can be written as:
\begin{align}
    \rho=\rho_{\chi}+\rho_{\scaleto{A}{4pt}}\; , \label{eq:rho_CN}
\end{align}
where
\begin{align}
    \rho_{\chi}=\frac{\dot{\chi}^2}{2}+\mu^4\left(1+\cos\frac{\chi}{f}\right)\;,\label{eq:rho_Chi}
\end{align}
and
\begin{align}
    \rho_{\scaleto{A}{4pt}}=\frac{1}{2\beta^4}\left(\dot{\psi}+\dot{\alpha}\psi -2\left(\dot{\sigma}+\frac{\dot{\beta}}{\beta}\right)\psi\right)^2 +\beta^2\left(\dot{\psi}+\dot{\alpha}\psi +\left(\dot{\sigma}+\frac{\dot{\beta}}{\beta}\right)\psi\right)^2 +g_{\scaleto{A}{4pt}}^2\frac{(2+\beta^4)\psi^4}{2\beta^2}\;.\label{eq:rho_A}
\end{align}
The isotropic pressure is then given by
\begin{align}
    P=\dot{\chi}^2 -\rho_{\chi}+\frac{\rho_{\scaleto{A}{4pt}}}{3}\;,\label{eq:P_isotropic}
\end{align}
while the anisotropic part is
\begin{align}
    \Tilde{P}=&\frac{1-\beta^6}{3}\left(\frac{1}{\beta^4}\left(\dot{\psi}+\dot{\alpha}\psi -2\left(\dot{\sigma}+\frac{\dot{\beta}}{\beta}\right)\psi\right)^2 -\frac{1}{\beta^2}g_{\scaleto{A}{4pt}}^2\psi^4\right)\label{eq:P_Anisotropic}\\ \nonumber
    & -\beta^2\left(\frac{\dot{\beta}}{\beta}+\dot{\sigma}\right)\left(2\dot{\psi}+2\dot{\alpha}\psi -\left(\dot{\sigma}+\frac{\dot{\beta}}{\beta}\right)\psi\right)\psi.
\end{align}
One constant of motion arising from this effective Lagrangian is the $\sigma$ associated momentum:
\begin{align}
    \Pi_{\sigma}=2 \beta^2 \psi \left[\psi\left(\dot{\sigma}+\dot{\alpha}+\frac{\dot{\beta}}{\beta}\right)+\dot{\psi}\right]-\frac{2 \psi}{\beta^4} \left[\psi \left(-2\left( \dot{\sigma}+\frac{\dot{\beta}}{\beta}\right)+\dot{\alpha}\right)+\dot{\psi }\right]+6 \dot{\sigma}\; ,
\end{align}
leading to:
\begin{align}
	\dot{\sigma}=\frac{\beta^4 De^{-3\alpha} - \left(\frac{\dot{\beta}}{\beta} \left(\beta^6+2\right)\psi^2 +\left(\dot{\psi}+\dot{\alpha}\psi\right)\left(\beta^6-1\right)\psi\right)}{3\beta^4+\left(\beta^6+2\right)\psi^2},\label{SIGMA}
\end{align} 
where $D$ is some arbitrary constant, and represents the initial anisotropy conjugate momentum. In a previous study \cite{Maleknejad:2013npa} this constant was set to $D=0$, effectively reducing the parameter space. As $D$ is exponentially suppressed by the number of efolds $\alpha$, it could be argued that the system should be indifferent to this initial parameter. However, the understanding of this parameter as akin to an initial anisotropy `push' restricts the initial value to be of order $D\sim H$.
To evaluate the restrictions on $D$ we first assume inflation takes place. Formally, in a Bianchi type I geometry this means that $\ddot{\alpha}>0$ and $\dot{\alpha}^2>\dot{\sigma}^2$. Examining the first slow-roll parameter we have:
\begin{align}
    \varepsilon\equiv -\frac{\ddot{\alpha}}{\dot{\alpha}^2}=\frac{\tfrac{\rho+P}{2}+3\dot{\sigma}^2}{\dot{\alpha}^2}\label{eq:epsilon_Axion_SU(2)}.
\end{align}
Defining $\varepsilon_{iso}\equiv \frac{3\left(\rho+P\right)}{2\rho}$, we rewrite this as:
\begin{align}
    \varepsilon=\frac{\rho\varepsilon_{iso}}{3\dot{\alpha}^2}+3\left(\frac{\dot{\sigma}}{\dot{\alpha}}\right)^2.
\end{align}
Taking the upper limit on $\dot{\sigma}$ by setting $\varepsilon_{iso}\gtrsim 0$, we have:
\begin{align}
    1>\varepsilon\gtrsim 3\left(\frac{\dot{\sigma}}{\dot{\alpha}}\right)^2. \label{eq:Dominant_sigmadot}
\end{align}
Since all other constituents of the term \ref{SIGMA} have to be small during slow-roll as argued in \cite{Maleknejad:2013npa}, and assuming we start at an initial efolding number of $\alpha=0$, if $D$ dominates the term \ref{SIGMA}, we have $\dot{\sigma}\sim D/3$. This yields
\begin{align}
    \varepsilon\gtrsim\frac{D^2}{3\dot{\alpha}^2}.
\end{align}
Thus it is formally possible for $\dot{\sigma}$ to momentarily achieve values comparable to
$\dot{\alpha}$. However, this high value should be attenuated almost immediately to a value more suited 
for a slow-roll inflation. Specifically in these systems it was shown that $\varepsilon\simeq (1-n_s)^2$ \cite{Adshead:2013qp,Dimastrogiovanni:2012ew}, with $n_s$ being the scalar spectral index. This should push $\varepsilon$ to be of the order of $10^{-3}$, so we would expect $|D| \ll \dot{\alpha}$. Since $D$ is multiplied by a factor of $e^{-3\alpha}$, it was thought that whatever initial value is assigned, $D$ is irrelevant. However, when $D>$ several $\times\;\dot{\alpha}$, the system becomes unstable and the inflationary solution becomes non-viable. This represents a level of tuning previously overlooked \cite{Maleknejad:2013npa}.\\\\
\subsection{Numerical analysis results}\label{sec:Numerical-Pure_CN}
\begin{figure}[!h]
    \centering
    \includegraphics[width=0.85\textwidth]{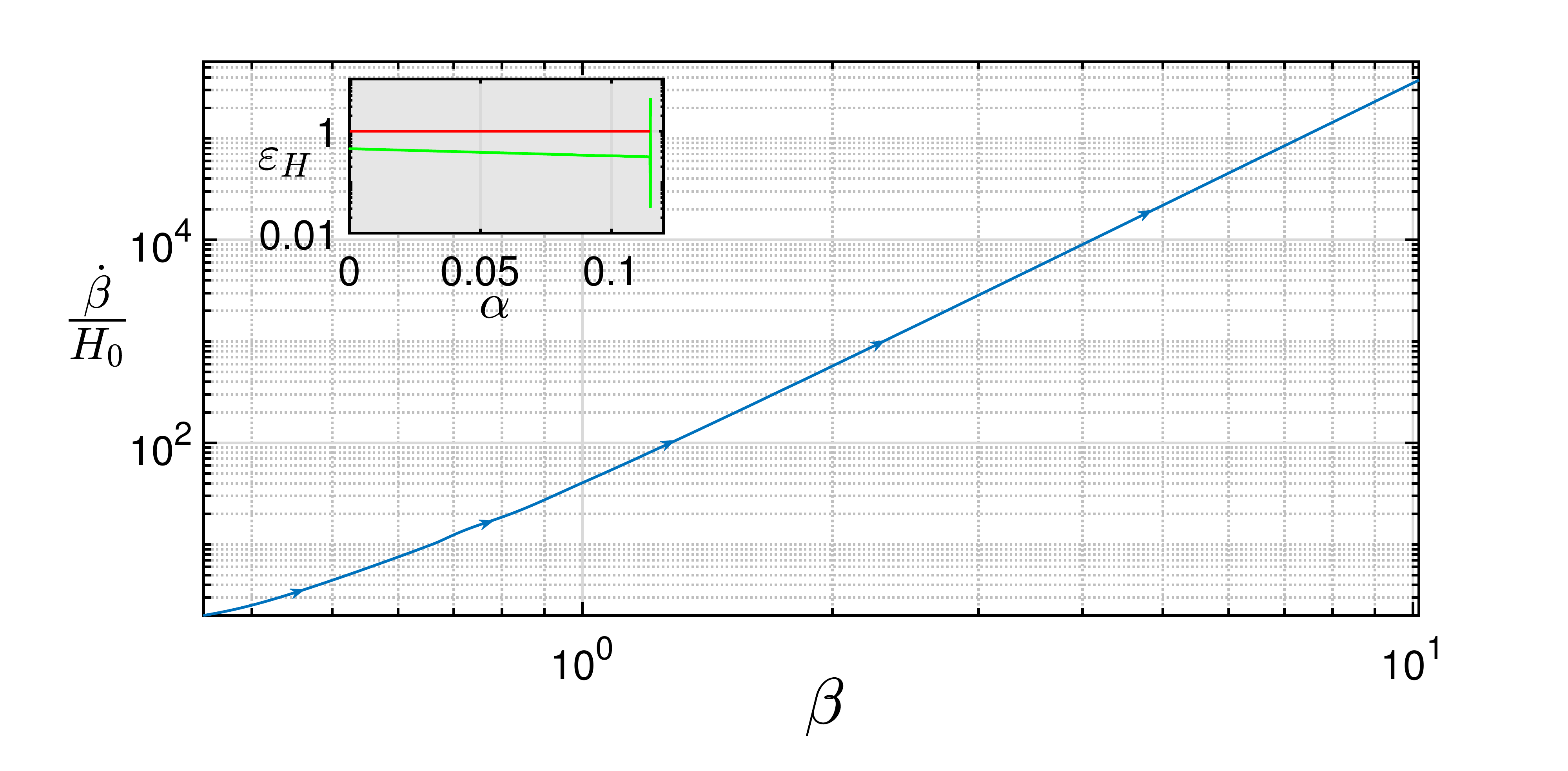}
    \caption{A trajectory that starts within the $\beta-\dot{\beta}$ phase space no-go region quickly evolves to values of $\frac{\dot{\beta}}{H_0 \beta}>10^5$ in $\sim 0.1$ efolds, before slow-roll inflation ends. The embedded graph shows the evolution of $\varepsilon_H$ as a function of efolds (green line). Slow-roll violation is evident when $\varepsilon_H$ climbs over 1 (red line) at approximately $\alpha\sim 0.12$.}
    \label{fig:beta_div_traj}
\end{figure}

\begin{figure}[!h]
    \centering
    \includegraphics[width=0.85\textwidth]{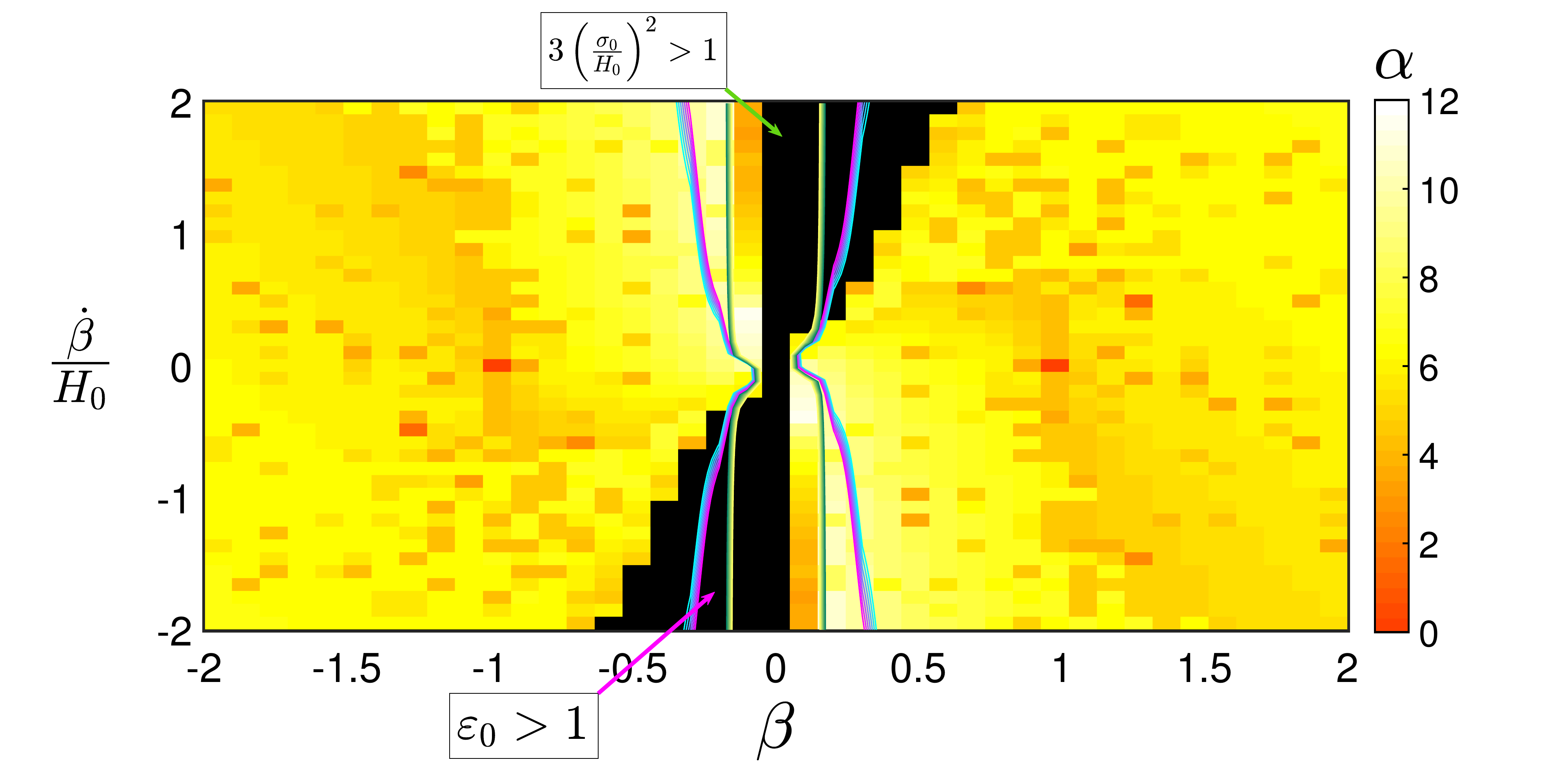}
    \caption{A convergence map of the CN system in the phase space of $(\beta,\dot{\beta}/H_0)\in([-2,2],[-2,2])$. 
    The map shows how many efolds it takes the system to fully isotropize. The black area is the no-go area where the system fails to isotropize. The area within the purple gradient denotes regions where the initial $\varepsilon_0>1$. The area within the blue-green gradient denotes regions where $3\left(\frac{\dot{\sigma}}{H_0}\right)^2>1$. We see the correspondence between no-go area and initial slow-roll violation is weak.}
    \label{fig:Slow_roll_no_go}
\end{figure}

Our numerical analysis uses the full equations of motion arising from the action \ref{eq:CN_action}. As mentioned before, the axion field is only coupled to the isotropic part of the gauge field, $\psi$. Therefore, the slow-roll relation \ref{eq:psi_FRW} is still valid in the Bianchi spacetime. We can thus use it to simplify the parameter space. We first specify at the onset of inflation $H_0$, $\chi_0,g_{\scaleto{A}{4pt}},\lambda$ and $f$. We derive $\mu$ using the slow-roll condition:
\begin{align}
    \mu=\left(\frac{3H_0}{1+\cos{\frac{\chi_0}{f}}}\right)^{1/4},\label{eq:Extract_mu}
\end{align}
and derive $\psi_0$ by:
\begin{align}
    \psi_0=\left(\frac{\mu^4\sin{\frac{\chi}{f}}}{3g_{\scaleto{A}{4pt}}\lambda H_0}\right)^{1/3}\; .\label{eq:Extract_psi}
\end{align}
Specifically, we set $H_0$ to the GUT scale $H_0=10^{-6}M_{Pl}$. We assign $\dot{\psi}_0=\dot{\chi}_0=0$ for simplicity. The baseline parameters we use are specified in Table~\ref{tab:Initial_parmeters}.
\begin{table}[!th]
    \centering
    \begin{tabular}{||c|c||c|c||}
\hline
     $H_0$&$10^{-6}$&$f$&0.1\\
     $\lambda$&2000&$g_{\scaleto{A}{4pt}}$&$2\cdot 10^{-6}$\\
     $\chi_0$& $\pi\cdot 10^{-3}$&$\dot{\chi}_0$&0\\
     $\dot{\psi}_0$&0& &\\
     \hline
     \hline
     \rowcolor{light-gray}
     $\mu$&$\sim 10^{-3}$&$\psi_0$&$\sim\pi/200$\\
     \hline
     \hline
     \rowcolor{light-gray}
     $\alpha$&$\gtrsim 70$& &\\
     \hline
\end{tabular}
    \caption{Baseline parameters for the numerical analysis of the CN system. The upper parameters: $H_0, f, \lambda, g_{\scaleto{A}{4pt}}$, and $\chi_0$ are specified. The parameters at the bottom are derived from the slow-roll conditions. Unless specifically studied, $D=0$. The slow-roll conditions are used only in the initial parameters derivation stage. These basic parameters along with $\beta>1,\;\dot{\beta}/H_0=0$ yield an e-folding number of at least 70.}
    \label{tab:Initial_parmeters}
\end{table}\\
\begin{figure}[!htbp]
\begin{center}
\includegraphics[width=0.85\textwidth]{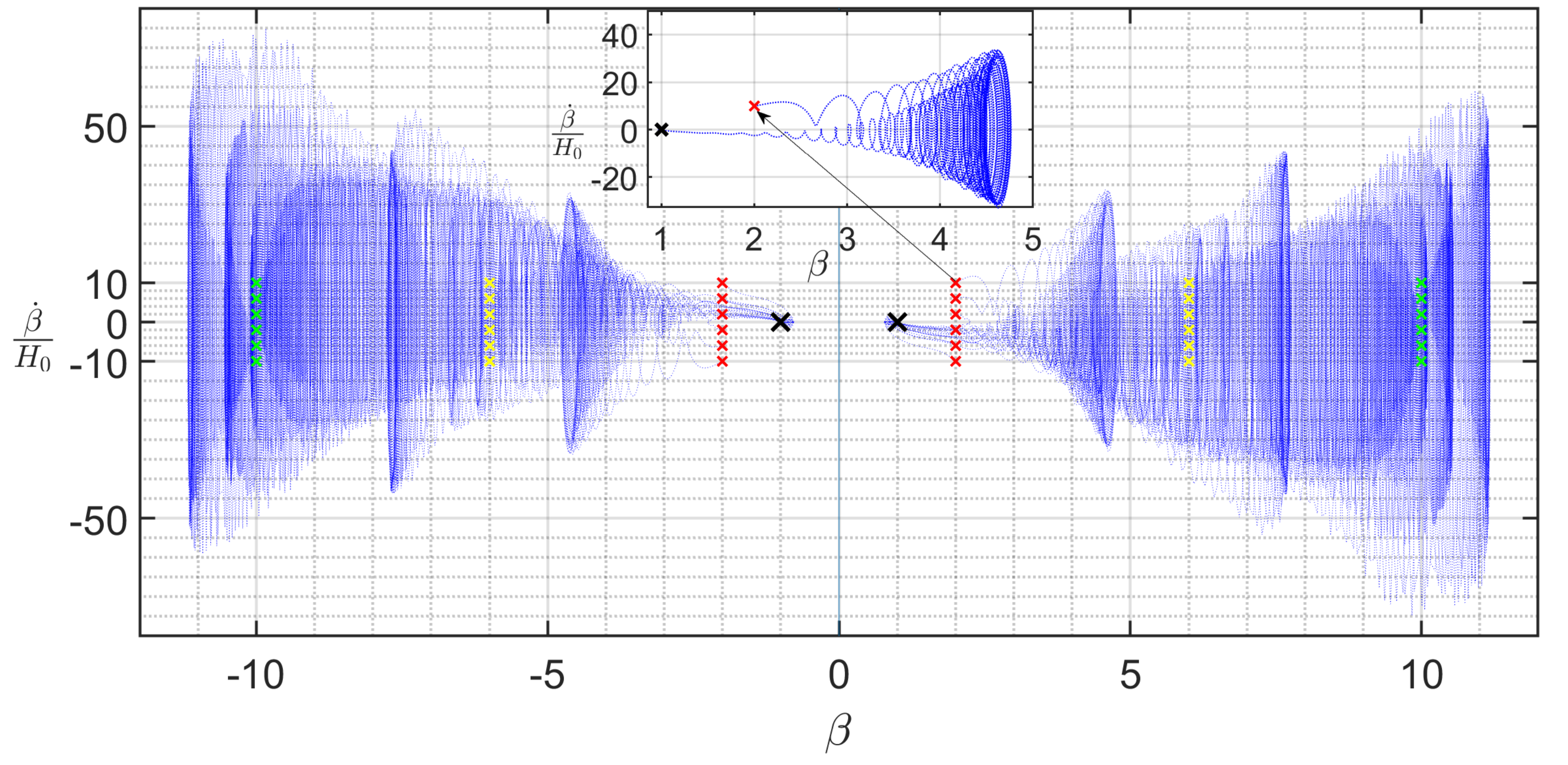}
\caption{Trajectories (blue dots) that start with various initial conditions (red,yellow,green X's) such that $(\beta_0,\frac{\dot{\beta}_0}{H_0})\in(\pm 10,\pm 10)$. Convergence to the attractors at $\beta=\pm 1$ (black X's) is apparent. While some trajectories initially turn away from the fixed points and anisotropy increases, they all eventually sufficiently isotropize. The embedded figure shows a single trajectory starting at $\beta_0=2,\frac{\dot{\beta}_0}{H_0}=10$.\label{fig:10x10original}}
\end{center}
\end{figure}
\begin{figure}[!htbp]
\begin{center}
\includegraphics[width=0.85\textwidth]{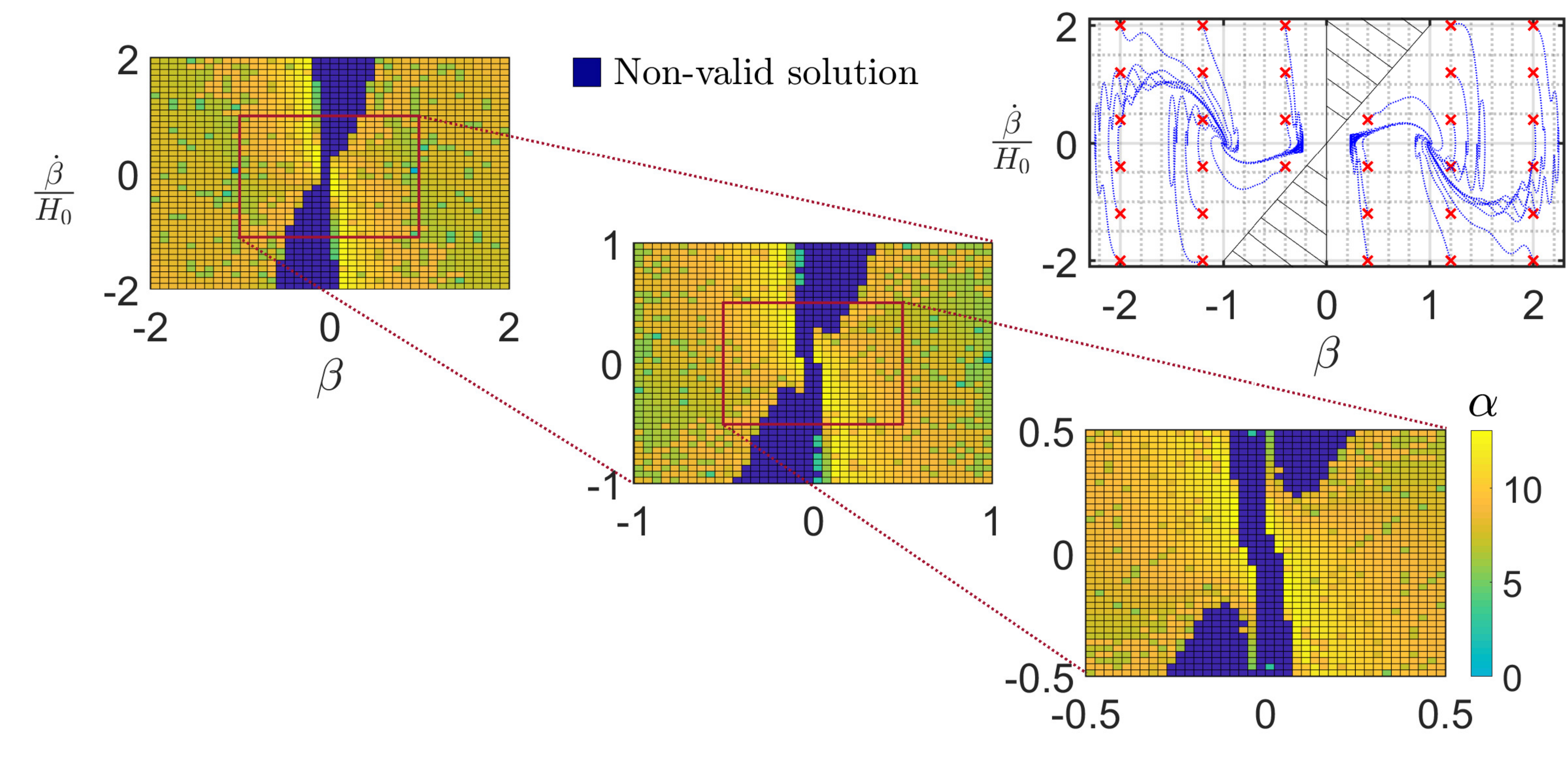}
\caption{A closer look at trajectories (upper right panel) in a smaller area of phase-space around $\beta=0$ reveals a `no-go' area (black striped area) where convergence is not achieved. This signifies that the system exits slow-roll ($\varepsilon_H>1$), and enters a `runaway' dynamics where the system cannot support inflation. The three panels from left to bottom right are convergence maps around $(0,0)$. The color shows the amount of efolds the system requires to isotropize. The system does not isotropize in the deep blue regions. Successive magnification around $(0,0)$ reveals an internal structure of this `no-go' area. \label{fig:Zoom_in}}
\end{center}
\end{figure}
The fixed point solution is given by $\beta=\pm 1$; however, it is hard to achieve this value numerically. Thus we employ effective convergence thresholds such that if $|\beta|=1\pm \Delta$, for a sufficiently small $\Delta$, with a vanishingly small $\dot{\beta}$ we are in the convergence target area. After enough consecutive integration steps inside the convergence area, we numerically set $\beta=\pm 1$ (corresponding to the correct attractor) and $\dot{\beta}=0$. \\
When calculating the number of efolds it takes the system to isotropize we use an additional exit condition of
\begin{align}
    \dot{\sigma}\ll \dot{\alpha}.
\end{align}
In practice, numerically we demand
\begin{align}
    |\dot{\sigma}(t)|<\dot{\alpha}(t)\cdot 10^{-3}.
\end{align}
We find that the isotropic attractor solutions at $\beta=\pm 1$ apparently hold for the full numerical analysis. Fig~\ref{fig:10x10original} shows a set of evenly spaced initial conditions at $(\beta,\dot{\beta}/H_0)\in\left(\pm 10, \pm 10\right)$. Trajectories starting at $\beta>0$ $\left(\beta<0\right)$ converge to $\beta=1$ $\left(\beta=-1\right)$, seemingly regardless of initial conditions. However, upon closer inspection we find some voids in the convergence pattern as seen in Fig.~\ref{fig:Zoom_in}. By gradually zooming into the area around the unstable point $(\beta,\dot{\beta}/H_0)=(0,0)$ we gain better resolution of these areas of no convergence. We call these `no-go' areas.\\
The existence of these areas can be understood by the looking at the gauge field kinetic terms of the system (Eq.~\ref{eq:rho_A}). We would expect that when $\frac{\dot{\beta}}{\dot{\alpha}\beta}>1$, the gauge field kinetic term dominates the energy density, and we quickly exit the slow-roll regime. In this case, while the gauge fields $\psi_i$ remain finite and $\beta\psi\ll 1$, taking the limit of $\frac{\dot{\beta}}{\dot{\alpha}\beta}\gg 1$ while setting $D=0$ for simplicity we have:
\begin{equation}
    \frac{\dot{\sigma}}{\dot{\alpha}}\simeq \frac{-\left[\frac{\dot{\beta}}{\dot{\alpha}\beta}\beta^6+\left(\frac{\dot{\psi}}{\dot{\alpha}\psi}+1\right)\beta^6\right]\psi^2}{3\beta^4}.
\end{equation}
Further, taking the limit where $\frac{\dot{\psi}}{\dot{\alpha}\psi}\ll 1$. we have:
\begin{equation}
    \frac{\dot{\sigma}}{\dot{\alpha}}\simeq - \left(\frac{\dot{\beta}}{\dot{\alpha}\beta}+1\right)\left(\beta\psi\right)^2.\label{eq:sigOverAlpha}
\end{equation}
Finally , again $\frac{\dot{\beta}}{\dot{\alpha}\beta}\gg1$ and so we arrive at:
\begin{equation}
    \frac{\dot{\sigma}}{\dot{\alpha}}\propto -\frac{\dot{\beta}\psi}{\dot{\alpha}}\beta\psi=-\frac{1}{\dot{\alpha}}\dot{\beta}\beta\psi^2
\end{equation}
Therefore, under these conditions the system will tend towards further anisotropization. Fig.~\ref{fig:Zoom_in} shows the areas of non-convergence around $\beta=0$. Surprisingly, some areas we should expect to fully anisotropize actually converge to the isotropic attractor. One may suspect that the no-go area coincides with the area of initial slow-roll violation ($\varepsilon_H>1$) or that of initial dominant anisotropy ($3\left(\frac{\dot{\sigma}}{\dot{\alpha}}\right)^2>1$, see equation~\ref{eq:Dominant_sigmadot}.) However, figure~\ref{fig:Slow_roll_no_go} shows this is not the case. \\
Interestingly, combining equations~\ref{eq:Dominant_sigmadot} and \ref{eq:sigOverAlpha} we have:
\begin{align}
    \frac{1}{3}>\frac{1}{\varepsilon}\left(\frac{\dot{\sigma}}{\dot{\alpha}}\right)^2>0.
\end{align}
Thus, as $\frac{\dot{\beta}}{\dot{\alpha}\beta}\rightarrow \infty$, and with a small non-zero  $\beta\psi$, $\varepsilon$ has to go to infinity to preserve this inequality. This means that at this limit, slow-roll must be violated and inflation has to become non-valid. This is shown in figure \ref{fig:beta_div_traj}, where for a trajectory starting at $\left(\beta_0=0.35,\frac{\dot{\beta}_0}{H_0}=2\right)$, as $\frac{\dot{\beta}}{H_0 \beta}$ is driven to values over $\sim 10^{5}$ the value of $\varepsilon_H$ rapidly increases until slow-roll is violated. \\\\

Although the no-go area includes $\frac{\dot{\beta}}{\dot{\alpha}\beta}\gg1$, where the kinetic term of the gauge-field sector is very large, and the system is far away from isotropy, it is much larger. More precisely, the area in the phase space regions with $\beta$ of the opposite sign as $\dot\beta$ (i.e. $\beta \dot\beta<0$) is well behaved, even where initially $\varepsilon>1$ or more drastically $3\left(\frac{\dot{\sigma}}{\dot{\alpha}}\right)^2>1$. However, in the phase space regions where $\beta$ has the same sign as $\dot{\beta}$ (i.e. $\beta \dot{\beta}>0$), the no-go area starts around $\frac{\dot{\beta}}{\dot{\alpha}\beta}>4$ and persists even well outside of the initial $\epsilon_0>1$ area (See figure~\ref{fig:Slow_roll_no_go}). The reason for the narrowness of the no-go area in the the  $\beta\dot{\beta}<0$ regions of the phase space is the due to the inability of the trajectories to cross the $\beta=0$ line and hence $\frac{\dot{\beta}}{\dot{\alpha}}$ must initially rapidly decrease. Such constraint does not hold for the  $\beta \dot\beta>0$ part of the phase space.

It is interesting to compare previous treatments of this system under slow-roll constraints \cite{Maleknejad:2013npa,Maleknejad:2011jr,Adshead:2018emn}, with our unconstrained analysis. The full separatrix of the chromo-natural system is embedded in a 6 dimensional space, i.e. $\left(\chi,\dot{\chi},\psi,\dot{\psi},\beta,\dot{\beta}\right)$. Upon enforcing slow-roll and after dropping all slow-roll suppressed terms, one can factor out $\psi$ and $\chi$. Thus the dynamical system is constrained to a 2 dimensional (reduced) phase space for a given $\chi_0,\dot{\chi}_0,\psi_0,\dot{\psi}_0$. A separatrix then emerges in the $\beta-\dot{\beta}/\dot{\alpha}$ phase-space. However, even starting from slow-roll inflation, there is a region in the unconstrained phase space that does not fulfill that expectation, i.e. the no-go area. In other words, the system in the no-go area spans the full phase space, and the time evolution of the $\psi$ and $\chi$ fields are so large that the system quickly gets out of the 2 dimensional reduced phase space and terminates inflation. 

That said, the divergent region looks to coincide with the separatrix mentioned.
The following figure \ref{fig:sep_Azadeh} shows this, albeit in a rough-sketch manner. It is also worth mentioning that the phase space presented in \cite{Adshead:2018emn}, while having similar features, describes a different system then the `pure' SU(2)-axion coupling we study in the no-inflaton limit, i.e. it is 'Higgsed'. \\
\begin{figure}[!h]
    \centering
    \includegraphics[width=0.85\textwidth]{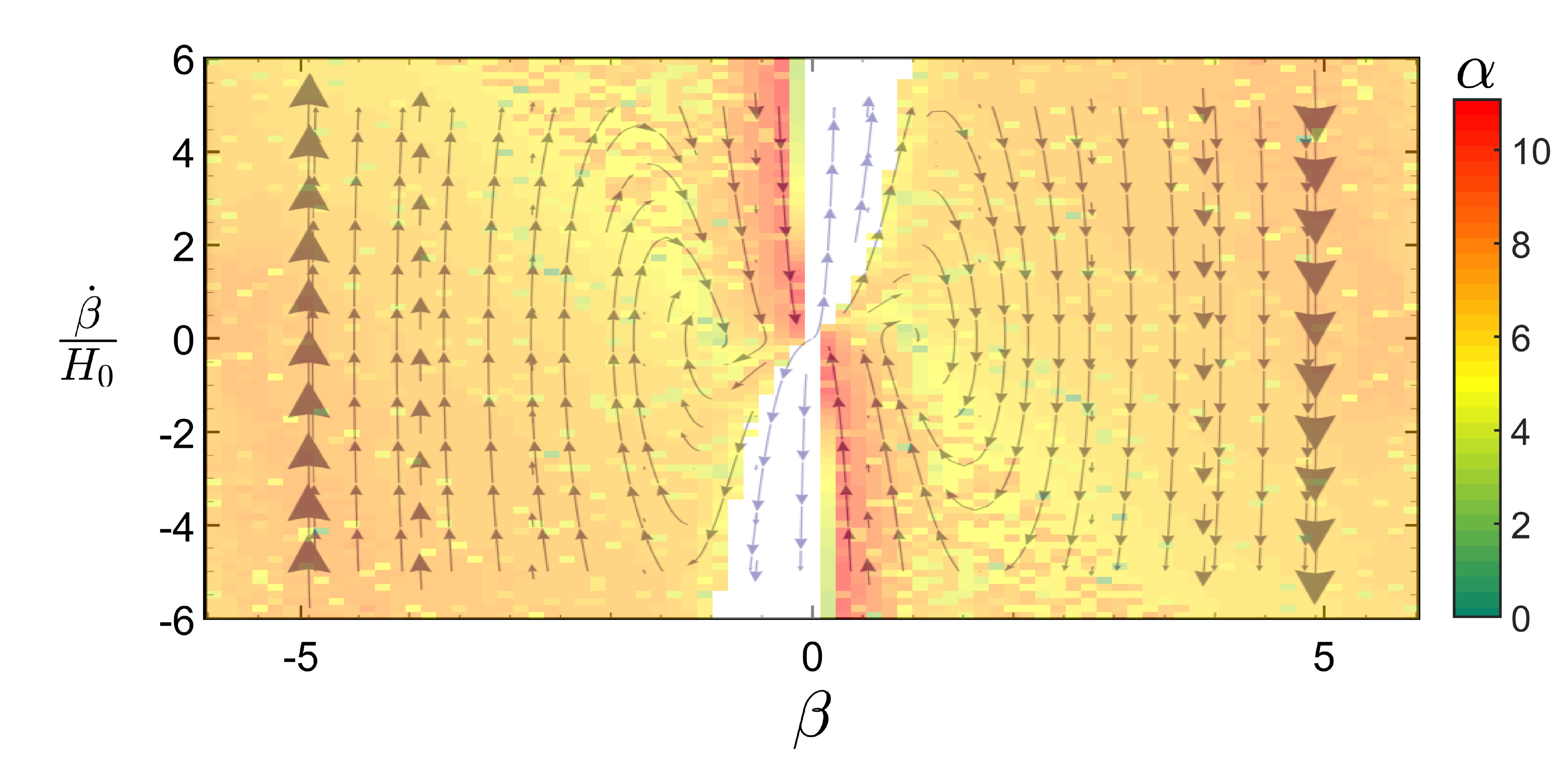}
    \caption{ An overlay of the convergence map over the phase-space flow diagram in \cite{Maleknejad:2013npa}. In the no-go region, the time evolution of fields ($\psi,\chi,\beta$) is very large which quickly terminates inflation. In other words, the slow-roll conditions assumed in earlier works to plot this 2D phase diagram is not valid in the no-go region.}
    \label{fig:sep_Azadeh}
\end{figure}
\newpage
\section{Spectator Axion-SU(2) Sector in Bianchi Type I Spacetime}\label{sec:Anisotropic-Spectator}
We now relegate the axion-SU(2) system to a spectator status, albeit embedded in a Bianchi type I geometry. Again due to the minimal coupling of the inflaton sector, the changes to the energy density and pressure are straightforward, as the inflaton field does not generate anisotropic pressure.
\begin{align}
    \rho=\rho_{\scaleto{CN}{4pt}}+\frac{\dot{\phi}^2}{2}+V(\phi)\; ,\\ \nonumber \\
    P= P_{\scaleto{CN}{4pt}}+\frac{\dot{\phi}^2}{2}-V(\phi)\; ,\\ \nonumber \\
    \Tilde{P}=\Tilde{P}_{\scaleto{A}{4pt}}\; ,
\end{align}
where $\rho_{\scaleto{CN}{4pt}},P_{\scaleto{CN}{4pt}},\Tilde{P}_{\scaleto{A}{4pt}}$ are given in \eqref{eq:rho_CN},\eqref{eq:P_isotropic},\eqref{eq:P_Anisotropic} respectively.
As in section \ref{sec:Spectator_FRW}, we now denote the energy density and pressure as:
\begin{align}
    \rho=\rho_{\phi}+\Delta\rho;\hspace{15pt}P=P_{\phi}+\Delta P.
\end{align}
With this notation we can evaluate the slow-roll parameter $\varepsilon$ in this system as:
\begin{align}
    \varepsilon=\varepsilon_{\phi}+\varepsilon_{\scaleto{CN}{4pt}}
\end{align}
where $\varepsilon_{\phi}=\frac{3}{2}\left(\frac{P_{\phi}+\rho_{\phi}}{\rho_{\phi}}\right)$, and $\varepsilon_{\scaleto{CN}{4pt}}=\left(\tfrac{\rho_{\scaleto{CN}{4pt}}+P_{\scaleto{CN}{4pt}}}{2}+3\dot{\sigma}^2\right)/\left(\tfrac{\rho_{\phi}}{3}+\dot{\sigma}^2\right)$. This evaluation is valid if, in addition to $\frac{\Delta \rho}{\rho_{\phi}},\frac{\Delta P}{P_{\phi}}\ll 1$ we also have $\frac{\dot{\sigma}}{\rho_{\phi}}\ll 1$. The second slow-roll parameter is  given by:
\begin{align}
    \eta=\frac{1}{\dot{\alpha}}\frac{\dddot{\alpha}}{\ddot{\alpha}}=\frac{1}{\dot{\alpha}}\frac{\partial}{\partial t}\ln{\ddot{\alpha}}.
\end{align}
Thus as in \eqref{eq:FRW_CN_eta} and \eqref{eq:eta_BIG_CN} we have two regions:
\begin{align}
    \left\{\begin{array}{llr}
         \eta\simeq&\eta_{\phi}+\sqrt{\frac{3}{\rho_{\phi}}}\frac{\partial}{\partial t}\left(\frac{\varepsilon_{\scaleto{CN}{4pt}}}{\varepsilon_{\phi}}\right)&\qquad \varepsilon_{\phi}\ll\varepsilon_{\scaleto{CN}{4pt}},  \\
         \eta\simeq&\eta_{\scaleto{CN}{4pt}}+\sqrt{\frac{3}{\rho_{\phi}}}\frac{\partial}{\partial t}\left(\frac{\varepsilon_{\phi}}{\varepsilon_{\scaleto{CN}{4pt}}}\right)&\qquad\varepsilon_{\scaleto{CN}{4pt}}\ll\varepsilon_{\phi},
    \end{array}\right.
\end{align}
where this time $\eta_{\scaleto{CN}{4pt}}\equiv\sqrt{\frac{3}{\rho_{\phi}}}\frac{\partial}{\partial t}\ln{(|\rho_{\scaleto{CN}{4pt}}+P_{\scaleto{CN}{4pt}}+6\dot{\sigma}^2|)}$, due to the different geometry.\\
We thus inherit the same analysis from the spectator system in FRW geometry regarding the stability conditions.
\subsection{Numerical results}
\begin{figure}[!htbp]
    \centering
    \includegraphics[width=0.85\textwidth]{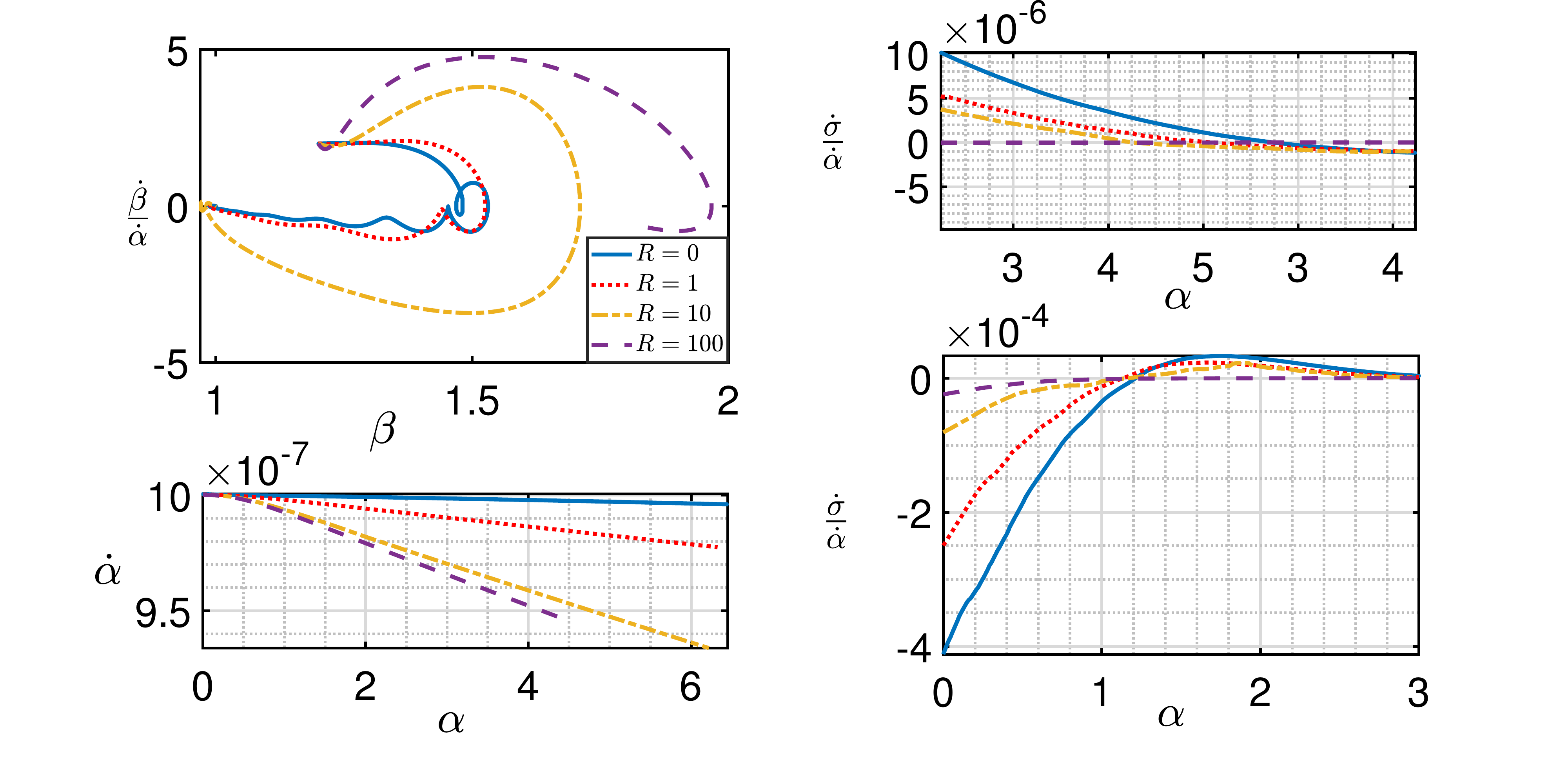}
    \caption{A comparison of trajectories and isotropization for identical intitial conditions, with different $R$ values. The upper left panel shows the trajectories in the $\beta-\frac{\dot{\beta}}{\dot{\alpha}}$ phase-space. The panel on the bottom right shows the $\dot{\alpha}$ evolution for each system. On the left hand the convergence of $\frac{\dot{\sigma}}{\dot{\alpha}}$ to the numerical threshold is shown. It is clear that as $R$ increases, the convergence is swifter. The $R=100$ plot (purple dash) achieves sufficient isotropization in under 4.5 efolds, and before the system converges on the attractor.}
    \label{fig:R-trajectories}
\end{figure}
\begin{figure}[!htbp]
\begin{center}
\includegraphics[width=0.85\textwidth]{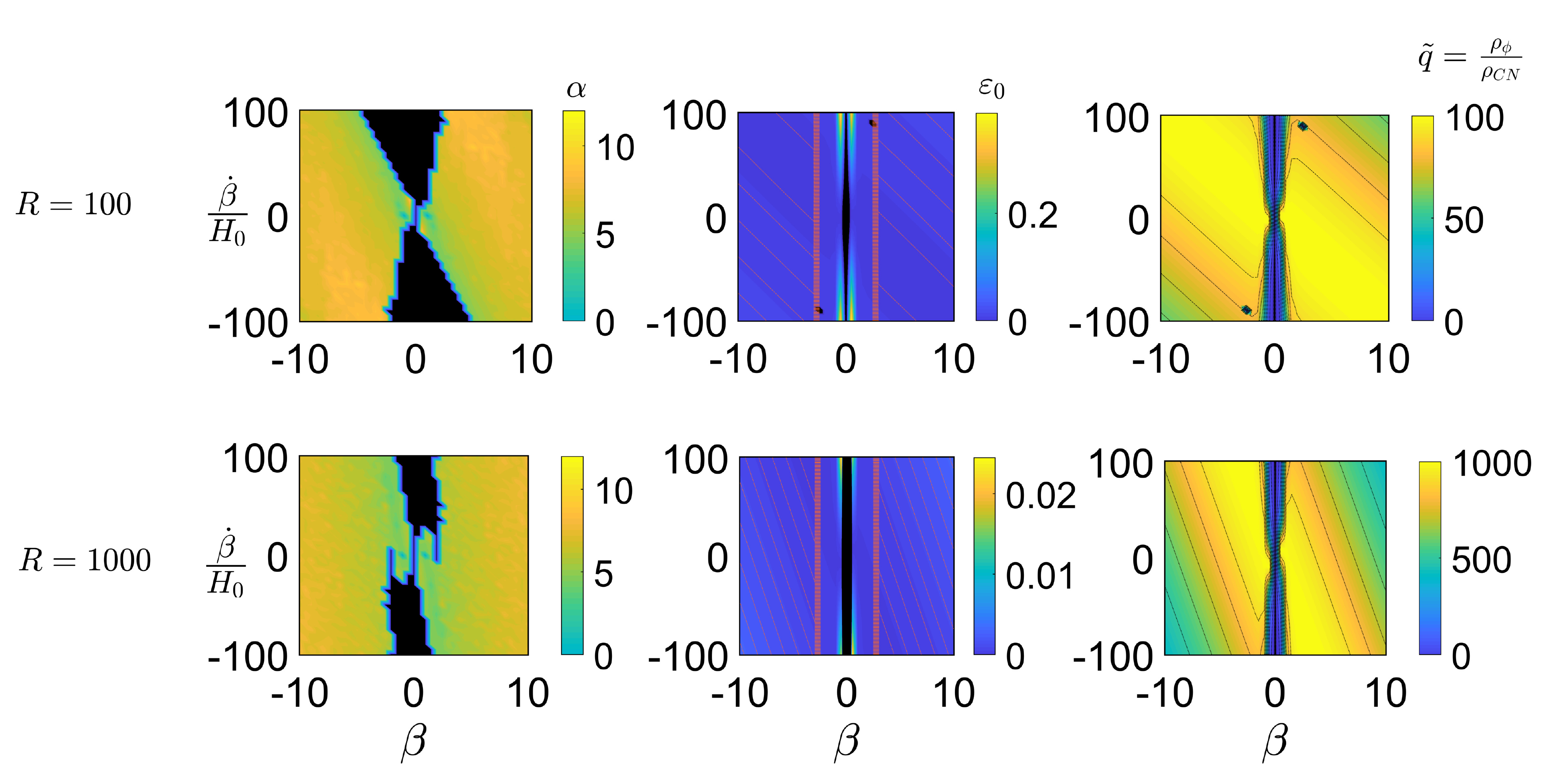}
\caption{Phase-space characteristics for $R=100$ (upper row) and $R=1000$ (lower row).
The columns from left to right are: convergence plot, initial $\varepsilon$ for each simulated system, and $\Tilde{q}$ for each system. On the second and third columns we also show the iso-contours of $\varepsilon$ (orange) and $\Tilde{q}$} (black). We see no correlation between the no-go area and areas of initial high $\varepsilon$ or low $\Tilde{q}$.\label{fig:Panels}
\end{center}
\end{figure}

We now use the action in \ref{eq:CNspect_action}, with an inflatonary potential that supports 70 efolds of inflation. We use the CN realization previously discussed, and ensure we have an overall GUT scale inflation by fixing $H_0=10^{-6}M_{Pl}$. We first define $R$ as the ratio between the inflationary and the axion potentials at the onset of inflation:
\begin{align}
    R\equiv\frac{V(\phi_0)}{\mu^4\left(1+\cos\frac{\chi_0}{f}\right)}.
\end{align}
Thus with $R,H_0=10^{-6},\chi_0=\pi\cdot 10^{-3},\phi_0=0$ and $f=0.1$, we can extract $\mu$ and $V(\phi_0)$ by :
\begin{align}
   \mu=\left(\frac{3H_0^2}{(1+R)\left(1+\cos\frac{\chi_0}{f}\right)}\right)^{1/4}\; , \\ \nonumber \\
   V(\phi_0)=R\mu^4\left(1+\cos\frac{\chi_0}{f}\right)\; .
\end{align}
 Additionally, we extract $\psi_0$ from \ref{eq:Extract_psi}, with $g_{\scaleto{A}{4pt}}=2\cdot 10^{-6}$ and $\lambda=2000$. The remaining kinetic terms are set to $\dot{\chi}_0=\dot{\psi}_0=0$. \\
Every simulated evolution starts with an initial $(\beta,\dot{\beta})$ pair. The overall Hubble parameter is set to $H_0=10^{-6}M_{Pl}$, but it is done in the absence of a kinetic term for $\beta$. Thus, we re-calculate the actual initial Hubble parameter $\dot{\alpha}_0$ including the $\beta,\dot{\beta}$ terms. We then solve the equations of motion numerically for each integration step, while taking care to use the Friedmann equation to check our result. This effectively is a consistency check on our calculations, and typically we have $\dot{\alpha}^2 -\dot{\sigma}^2 -\frac{\rho}{3}=\mathcal{O}(10^{-27})$, which is at the level of machine precision, for the smallest fixed quantity in the system - $\mu^4$.\\
\begin{figure}[!htbp]
\begin{center}
\includegraphics[width=0.85\textwidth]{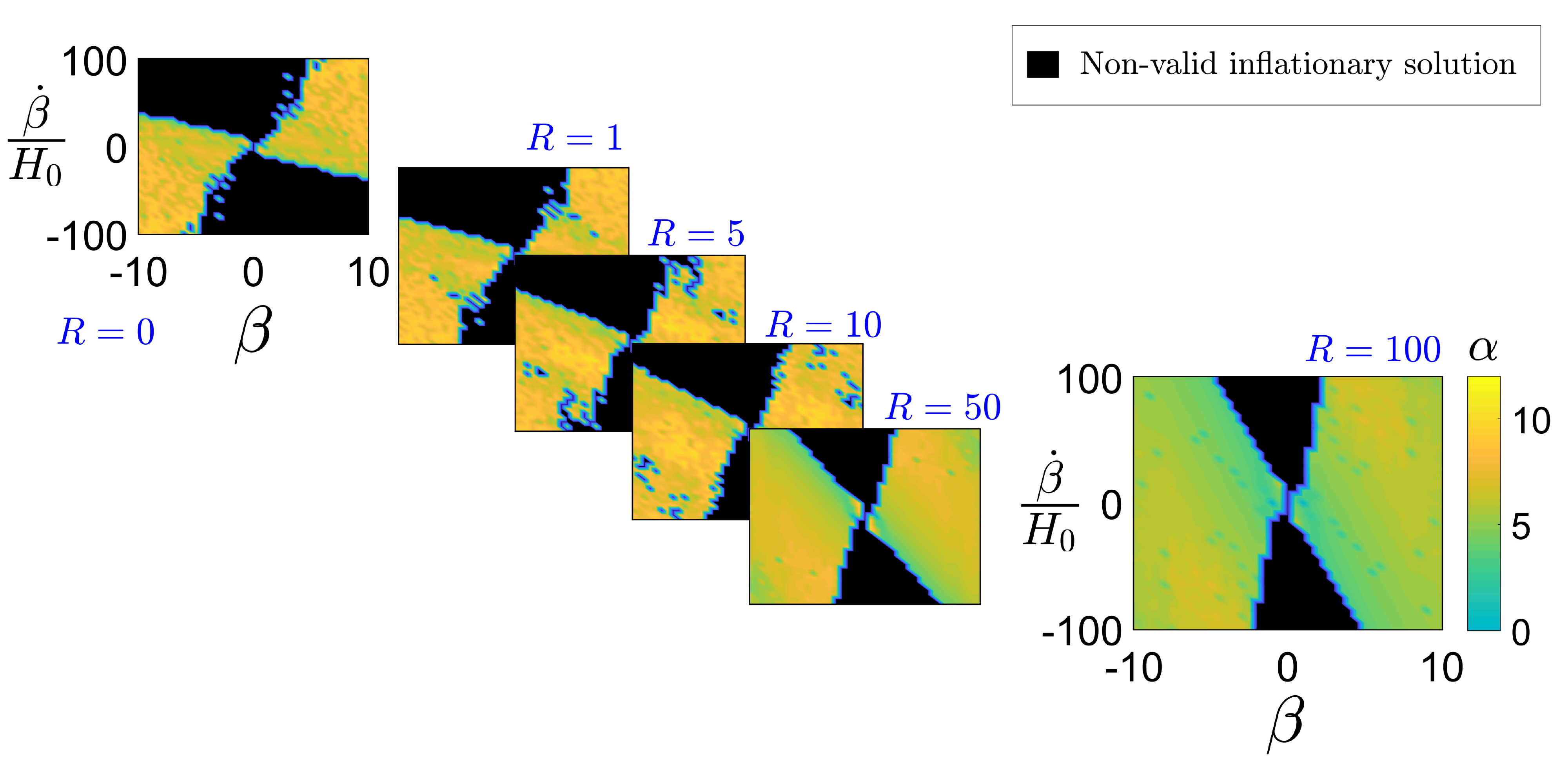}
\caption{Gradually increasing the  inflaton energy density with respect to the spectator axion energy density. The parameter $R$ encodes the domination of the inflaton component over the axion component. We probe the larger phase-space of $(\beta,\dot{\beta}/H_0)\in(10,100)$ to look at the limits of the basin of attractor. The phase-space convergence plots show that as the inflaton becomes more dominant, the system becomes indifferent to initial conditions, and readily converges to the isotropic attractor solution. The color bar shows the number of efolds it takes the system to isotropize.}\label{fig:TurningOnInflaton}
\end{center}
\end{figure}\\
We employ the same numerical methods and exit conditions as in section \ref{sec:Numerical-Pure_CN}. However, when one wishes to compare systems with different $R$ values, a fixed threshold is needed; thus we use
\begin{align}
   |\dot{\sigma}(t)|<H_0\cdot 10^{-9}=10^{-15}, 
\end{align}
where $H_0$ is fixed regardless of $R$, and $\dot{\alpha}(t)$ is time-dependent and sensitive to the system's specific setup.
We first look at a few trajectories, studying the effect of an increase of $R$. In figure~\ref{fig:R-trajectories}, we observe that as $R$ increases, trajectories tend to stabilize (top left panel) in the sense of less oscillations in the $\beta-\frac{\dot{\beta}}{\dot{\alpha}}$ phase-space. This is due to the dominance of $V(\phi)$ over the axion potential scale $\mu$. We also observe that the system isotropizes more quickly (bottom and top right panel) with the increase of $R$.\\
\begin{figure}[!htbp]
\begin{center}
\includegraphics[width=0.85\textwidth]{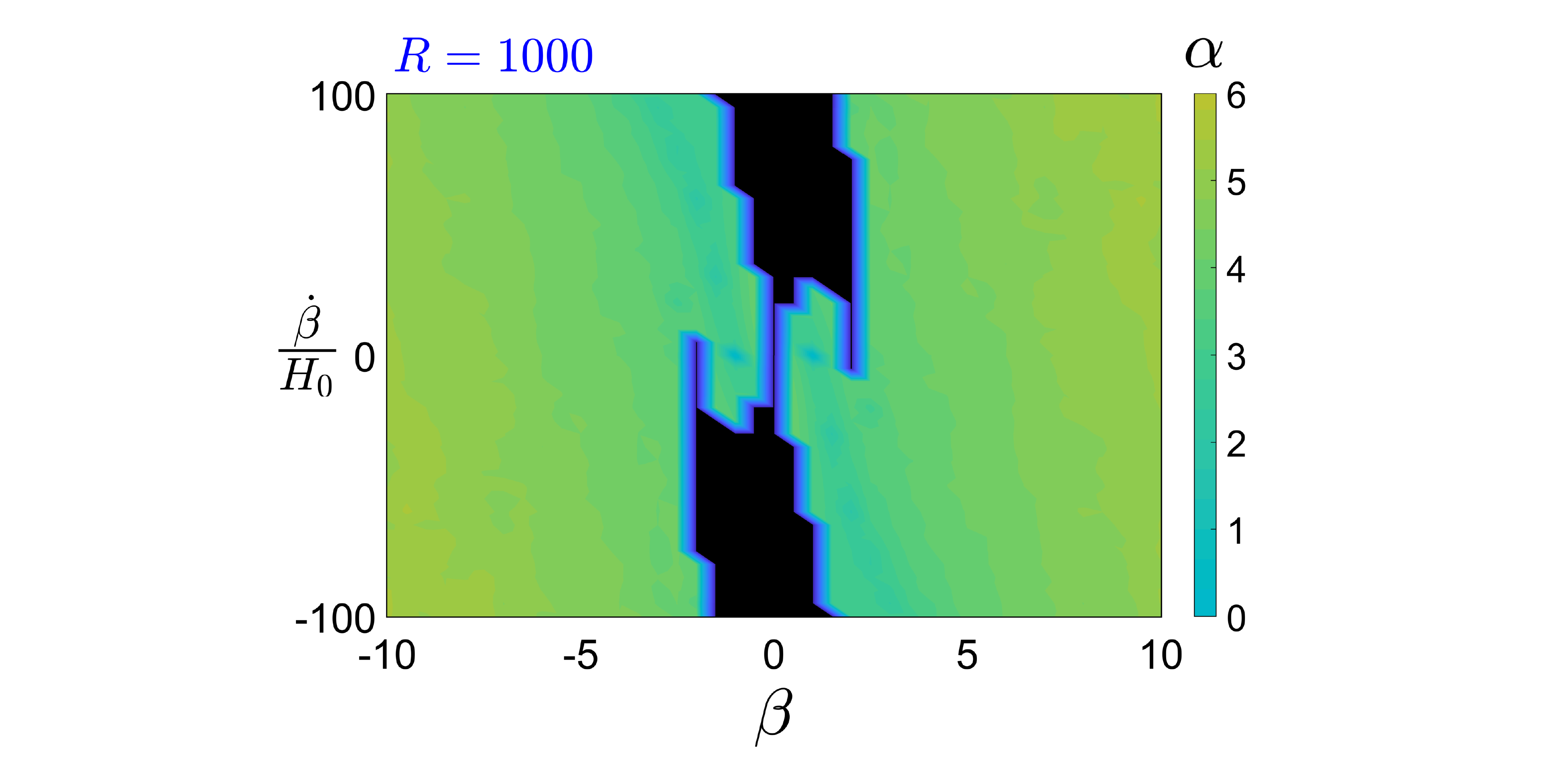}
\caption{Anisotropy phase space for $R=1000$. While we see that all regions that converge to the attractor solution do so in $\sim 6$ efolds, we still observe the persistence of the no-go areas. This is due to the kinetic term `hand-off' mechanism.}\label{fig:R1000}
\end{center}
\end{figure}\\
We looked at a larger phase-space than previously studied to probe the limits of the basin of attraction. By simulating anisotropy phase-spaces with different $R$ values, we find that the system gradually stabilizes into an inflaton dominated one ,i.e., the usual scalar-driven slow-roll inflation. Fig. \ref{fig:TurningOnInflaton} shows this, as even the larger phase-space, where $(\beta_0,\frac{\dot{\beta}_0}{H_0})\in (\pm 10,\pm 100)$ supports the attractor isotropic solution.\\
Even so, there are still locations in the anisotropy phase-space where initial conditions violate slow-roll inflation. This is due to the gauge field kinetic energy term dependence on $\beta$. Recall that $\beta=0$ is an unstable fixed point for the $\psi$ and $\beta$ field equations. Sufficiently close to $\beta=0$ the kinetic term in equation~\ref{eq:rho_A} part that is proportional to $\frac{\dot{\beta}^2}{\beta^6}$ becomes unbounded, violates slow-roll, and drives the system to large values of $\beta$. At that point there are two options. Either the system stabilizes and converges to $\beta=\pm 1$ albeit slowly, or the large value of $\dot{\beta}$ is `handed-off' to the kinetic term in equation~\ref{eq:rho_A} that is proportional to $\dot{\beta}^2$. In turn this feeds the anisotropy momentum, such that in the limit of large $\frac{\dot{\beta}}{\dot{\alpha}\beta}$ we have:
\begin{align}
    \left\{\begin{array}{llc}
         \frac{\dot{\sigma}}{\dot{\alpha}}\simeq&-\frac{1}{3}\left(1+\tfrac{\dot{\beta}}{\beta\dot{\alpha}}\right)\left(\beta\psi\right)^2&\qquad \beta\gg1 \\
         \frac{\dot{\sigma}}{\dot{\alpha}}\simeq&\frac{1}{3}\left(1-\tfrac{2\dot{\beta}}{\beta\dot{\alpha}}\right)\left(\frac{\psi}{\beta^2}\right)^2&\qquad \beta\ll1
    \end{array}\right.
\end{align}
The energy density in the gauge-field sector goes like
\begin{align}
    \rho_{\scaleto{A}{4pt}}\sim \left( \frac{\dot{\beta}}{\beta}\right)^2\left(\frac{2}{\beta^4}+\beta^2\right)\psi^2=\left( \frac{\dot{\beta}}{\beta}\right)^2\left(2\left(\frac{\psi}{\beta^2}\right)^2+\left(\beta\psi\right)^2\right)
\end{align}
Therefore, in the regions where either $\beta\rightarrow 0$, or $\frac{\dot{\beta}}{\dot{\alpha}\beta}\gg1$ the system becomes mathematically unbounded. Note that $\beta= 0$ is a singularity of paramtrization. However, all of the physical quantities, i.e. $\beta\psi$ and $\frac{\psi}{\beta^2}$ are always finite. Therefore, the physical system with $\beta=0$ implies $\beta\psi=\frac{\psi}{\beta^2}=0$, i.e. zero gauge fields VEV.  Adding the inflaton potential may go a long way to stabilize these trajectories, but ultimately, these no-go areas should not be expected to vanish altogether. We see that even at $R=1000$, i.e., the system is highly dominated by the inflationary potential, these no-go areas still persist when $\dot{\beta}/H_0\gg 1$. This area has not been previously studied. This is shown in figure~\ref{fig:R1000}.\\\\
One may suspect these areas coincide with areas where the energy density in the CN sector, including the kinetic terms, initially dominates the inflaton sector. We define $\Tilde{q}$ as the ratio between the energy densities such that:
\begin{align}
    \Tilde{q}=\frac{\rho_{\phi}}{\rho_{\scaleto{CN}{4pt}}},
\end{align}
where the energy densities now contain all relevant kinetic terms. In figure~\ref{fig:Panels} we show that the no-go areas do not fully correlate to areas where the CN energy density dominates, even in the presence of kinetic terms. In other words, the no-go area is a dynamical effect due to the instability of the system around $\beta\sim 0$ and $\frac{\dot{\beta}}{\dot{\alpha}\beta}\gg1$.
\\
Moreover, except along a narrow strip along the $\beta\sim 0$ region, the system starts at slow-roll but fails to isotropize, even when the energy in the inflaton sector greatly dominates that of the CN sector. 
\section{Conclusion}\label{sec:Conclusion}
We have studied an axion-SU(2) gauge field system embedded in a homogeneous but anisotropic Bianchi type I geometry. In both of our realizations of the CN system \cite{Adshead:2012kp,Dimastrogiovanni:2016fuu} we parametrize the gauge field deviation from isotropy by the dimensionless parameter $\beta(t)$ \cite{Maleknejad:2013npa}. In this parametrization the prefect isotropic state is given by $\beta=\pm 1$, and a singularity exists at $\beta=0$. We have solved these systems numerically, using the full equations of motion and an energy consistency condition. We have investigated the attractor's basin of attraction to the homogeneous isotropic solution and confirmed that the system has a single attractor solution in each half-plane (corresponding to $\beta=\pm 1$). \\\\
We have found that, contrary to previous claims based on its exponential suppression, the initial value for $D$ cannot exceed several times the initial Hubble parameter $\dot{\alpha}$. When $D>$ a few $\dot{\alpha}$ the system exits slow-roll either directly, or through the `hand-off' mechanism.\\\\ 
We additionally explored a part of the $\dot{\beta}-\beta$ phase space which has not been previously studied. This regime, e.g. the vicinity of $\beta=0$ with a large kinetic term ($\dot{\beta}/H \gg 1$), contains a no-go area in which the anisotropies do not isotropize and the large kinetic energy in the anisotropies terminates inflation after a few efolds at the most. Thus these areas fail to support the necessary 60 or so efolds of inflation required to account for observations.\\\\
In the case of the spectator axion-SU(2) system \cite{Dimastrogiovanni:2016fuu}, stability is significantly increased. The anisotropies decay faster and the system settles into the attractor solution in fewer efolds. The basin of attraction can additionally support larger kinetic terms, and the no-go areas substantially shrink.\\

Therefore we conclude that in terms of tuning requirements, and in the sense of phase-space stability the spectator axion-SU(2) model is more viable. Thus it is phenomenologically more attractive as a  methodological substrate, rather than the pure CN system, to study inflationary models which involve primordial SU(2) symmetry.

\section*{Acknowledgements}
This research was supported in part by the Excellence Cluster ORIGINS which is funded by the Deutsche Forschungsgemeinschaft (DFG, German Research Foundation) under Germany's Excellence Strategy – EXC-2094 – 390783311.\\
EK thanks Shinji Mukohyama for the question he asked during the `General Relativity - The Next Generation' conference held at Yukawa Institute for Theoretical Physics from February 19th to 23rd, 2018, which led to this project.\\
IW thanks the Minerva Foundation and the Max Planck Society, for supporting his research.

\end{document}